\documentclass[12pt,preprint]{aastex}

\def\lesssim{\mathrel{\hbox{\rlap{\hbox{\lower4pt\hbox{$\sim$}}}\hbox{$<$}}}}
\def\gtrsim{\mathrel{\hbox{\rlap{\hbox{\lower4pt\hbox{$\sim$}}}\hbox{$>$}}}}

\slugcomment{submitted to ApJ.}

\shorttitle{Collapse of Rotating Magnetized Molecular Cloud Core and Mass Outflows}
\shortauthors{K. Tomisaka}

\begin{document}

\title{Collapse of Rotating Magnetized Molecular Cloud Cores and Mass Outflows}

\author{Kohji Tomisaka}
\affil{Theoretical Astrophysics, National Astronomical Observatory,
Mitaka, Tokyo 181-8588, Japan}
\email{tomisaka@th.nao.ac.jp}

\begin{abstract}
Collapse of the rotating magnetized molecular cloud core is studied
 with the axisymmetric magnetohydrodynamical (MHD) simulations.
Due to the change of the equation of state of the interstellar gas,
 the molecular cloud cores experience several different phases as collapse proceeds.
In the isothermal run-away collapse ($n \lesssim 10^{10}{\rm H_2\,cm}^{-3}$),
 a pseudo-disk is formed and it continues to contract till the opaque core is formed at the center.
In this disk, a number of MHD fast and slow shock pairs appear 
 running parallelly to the disk.
After the equation of state becomes hard, an adiabatic core is formed, which is separated
 from the isothermal contracting pseudo-disk by the accretion shock front facing radially outwards. 
By the effect of the magnetic tension, the angular momentum is transferred from the disk mid-plane
 to the surface.
The gas with excess angular momentum near the surface is finally ejected, which explains the molecular bipolar outflow.
Two types of outflows are observed.
When the poloidal magnetic field is strong (magnetic energy is comparable to the thermal one),
 a U-shaped outflow is formed 
 in which fast moving gas is confined to the wall whose shape looks like a capital letter U.
The other is the turbulent outflow in which magnetic field lines and velocity fields are randomly oriented.   
In this case, turbulent gas moves out almost perpendicularly from the disk.  
The continuous mass accretion leads to the quasistatic contraction of the first core.
A second collapse due to dissociation of H$_2$ in the first core follows.
Finally another quasistatic core is again formed by atomic hydrogen (the second core).
It is found that another outflow is ejected around the second atomic core, which seems to correspond to
 the optical jets or the fast neutral winds.
\end{abstract}


\section{Introduction}

Star formation has been a long-standing target in astrophysics.
The infrared protostar distribution revealed that the
 molecular cloud cores, which coincide with relatively high-density part 
($n\sim 10^4{\rm cm}^{-3}$) of the molecular clouds,
 are the sites of star formation.
The observed molecular cloud cores are divided into two categories:
 those observed associated with and without protostars.
The molecular cloud cores without protostars are called
 starless cores or prestellar cores and are considered 
 younger than the cores associated with protostars (protostellar cores).
From a theoretical point of view, 
 clouds or cloud cores experience the isothermal run-away collapse first 
 and then accretion on to the stellar core develops \citep{lar69}.  
Prestellar cores which indicates inflow motions (e.g. in L1544 
 rotation and infall velocities $\sim 0.1 {\rm km\,s^{-1}}$ are
 observed by \citet{oha00}) do
 indicate that they are in the dynamically contracting phase, in other words,
 in the run-away collapse \citep{cio00}.
After the epoch when the dust thermal emissions are trapped in the
 central part of the cloud ($n_c \sim 10^{10}{\rm cm^{-3}}$),
 an adiabatic core is formed and isothermal gas continues to accrete
 on to the core.
The molecular cloud core in this phase is observed as a protostellar core.
It is shown that the dynamical evolution of the cloud core is
 characterized by the sequence from the prestellar cores to protostellar cores.

Dynamical collapse of the magnetized clouds are studied by many
 authors \citep{sco80,phi86a,phi86b,dor82,dor89,ben84,mos91,mos92,fie92,fie93,
 bas94,tom95,tom96,nak99}.
Rotating clouds collapse has been attacked seriously with numerical
 simulations \citep{bod80,nor80,woo82,nar84,tru97,tru98,tsu99a,tsu99b}. 
However, a restricted number of articles are published regarding
 the dynamical contraction of the cloud with both rotation and magnetic fields
 \citep{dor82,dor89,bas94,bas95};
 for quasistatic evolution see \citet{tom90}.
These researches are confined to the relatively early prestellar stage. 

Is it sufficient to consider the effects of the rotational motion and
 the magnetic fields separately?
In the dynamical contraction phase, it is shown that 
 the molecular outflow is driven by the cooperative effect of the
 magnetic fields and rotation motion \citep{tom98}.  
The toroidal magnetic fields are generated from the poloidal ones by
 the effect of rotation motions.
The magnetic torque works only when the poloidal and toroidal
 magnetic fields coexist.
The magnetic torque and thus magnetic angular momentum transfer
 along the magnetic field line is important to eject the outflow.
Since the outflow brings the excess angular momentum, the angular 
 momentum that remain in the adiabatic core and thus new-born star is
 reduced by a factor from $10^{-2}$ to $10^{-3}$ from that of the
 parent molecular cloud core \citep{tom00}.
The outflow have not observed either 
 in the magnetized but non-rotating cloud \citep{sco80,tom96}
 or the rotating but non-magnetized cloud \citep{nor80}.
Therefore, rotation and magnetic fields are both essential to the 
 evolution of molecular cloud cores.
In the present paper, we present the dynamical contraction
 of the magnetized and rotating cloud.
  
The cooperative effect of magnetic fields and the rotation motions 
 becomes important after the adiabatic core is formed near the center
 of the cloud core \citep{tom98}.
Therefore, the evolution throughout from the prestellar to
 protostellar core should be studied.

Plan of this paper is as follows: in $\S$2 model and numerical method are described.
As the initial condition, we choose a slowly rotating cloud with purely poloidal magnetic fields
 (no toroidal magnetic fields).
And we follow the evolution using magnetohydrodynamical simulations.
Section 3 is devoted to the numerical results.
In this section we compare clouds with strong magnetic fields and those with weak magnetic fields. 
This shows that completely different two types of outflows are ejected in respective clouds.
Another comparison is made between fast rotators and slow rotators.
In $\S$4, we discuss the evolution till the second core, which becomes actually a new-born star,
 is formed.
It is found that another outflow is found around the second core, which seems to correspond to
 the optical jets or high speed neutral winds.  
We also discuss whether the mass inflow/outflow rate and the momentum outflow rate
 observed in molecular bipolar outflows are explained or not. 
 
\section{Model and Numerical Method}

To study the dynamical contraction, we consider an isothermal cylindrical cloud
 in hydrostatic balance with infinite length as the initial state.
In terms of the gravitational potential $\psi$
 and isothermal sound speed $c_s$,
 the radial distributions of density $\rho$, rotation speed $v_\phi$,
 and magnetic field density $B_z$
 are calculated using the equation of hydrostatic balance as
\begin{equation}
  \frac{v_\phi^2}{r}
    -\frac{\partial \psi}{\partial r}
    -\frac{c_s^2}{\rho}\frac{\partial \rho}{\partial r}
    -\frac{1}{8\pi\rho}\frac{\partial B_z^2}{\partial r}=0,
							\label{eq1}
\end{equation}
and the Poisson equation for the self-gravity as
\begin{equation}
\frac{1}{r}\frac{\partial}{\partial r}
  \left(r\frac{\partial \psi}{\partial r}\right)=4\pi G \rho,
							\label{eq2}
\end{equation}
where $G$ represents the gravitational constant.
This equation has a solution as follows \citep{sto63}
\begin{equation}
  \rho_0(r)=\rho_{c\,0} \left( 1+ \frac{r^2}{8H_c^2} \right)^{-2},
							\label{eq3}
\end{equation}
\begin{equation}
  v_{\phi(r)\,0}\equiv r\Omega_0(r) 
  = r\Omega_{c\,0}\left(\frac{\rho}{\rho_c}\right)^{1/4} 
  = r \Omega_{c\,0} \left( 1+ \frac{r^2}{8H_c^2} \right)^{-1/2},
							\label{eq4}
\end{equation}
and
\begin{equation}
  B_{z\,0}(r)= B_{c\,0} \left(\frac{\rho}{\rho_c}\right)^{1/2}
        =B_{c\,0} \left( 1+ \frac{r^2}{8H_c^2} \right)^{-1},
							\label{eq5}
\end{equation}
with using the scale-height at the center, 
\begin{equation}
  H_c^2=\frac{c_s^2+B_{c\,0}^2/8\pi \rho_{c\,0}}
           {4\pi G \rho_{c\,0} - 2 \Omega_{c\,0}^2},
							\label{eq6}
\end{equation}
where $\rho_{c\,0}$, $\Omega_{c\,0}$, and $B_{c\,0}$ represent respectively
 the density, the angular rotation speed, and the magnetic flux density
 at the center of the cylindrical cloud ($r=0$).
This is the same hydrostatic configuration studied by \citet{mat97}.
However, we assume no initial toroidal magnetic field $B_\phi=0$ in
 contrast to them.
The density distribution extends till the radius where the thermal pressure
 $c_s^2 \rho_s$ becomes equal to the external pressure $p_{\rm ext}$.
The solution contains three non-dimensional parameters characterizing the
 distribution after adopting natural normalization such as for distance 
 $r' \equiv r/H \equiv r / [c_s/(4 \pi G \rho_s)^{1/2}]$,
 time  $t' \equiv t/\tau_{\rm ff} \equiv t / [1/(4 \pi G \rho_s)^{1/2}]$,
 and density $\rho' \equiv \rho/\rho_s \equiv c_s^2\rho(r)/p_{\rm ext}$.
We summarized the conversion factors from non-dimensional to physical
 quantities in Table \ref{table1}.
The first parameter characterizing the initial state is related to
 the magnetic to thermal pressure ratio as
\begin{equation}
  \alpha \equiv B_z(r)^2/4\pi\rho(r)c_s^2=B_{c\,0}^2/4\pi\rho_{c\,0} c_s^2
							\label{eq7}
\end{equation}
 and the second one is related to the angular rotation speed as
\begin{equation}
  \Omega'=\Omega_{c\,0}/(4\pi G \rho_s)^{1/2}.
							\label{eq8}
\end{equation}
Finally the surface to center density ratio 
\begin{equation}
  F\equiv \rho_{c\,0}/\rho_s.
							\label{eq9}
\end{equation}
The scale-height at the center $H_c$ is expressed using these
 non-dimensional parameters as
\begin{equation}
  H_c'^2=\frac{H_c^2}{c_s^2/4\pi G \rho_s}
      =\frac{1+\alpha/2}{F-2\Omega'^2}.  
							\label{eq10}
\end{equation}
From this equation,
 it is shown that a hydrostatic balance is achieved only when 
 $F > 2\Omega'^2$, i.e., $4\pi G \rho_{c\,0} > 2\Omega_{c\,0}^2$.
\citet{mat94}'s $\beta$ parameter, which represents the ratio
 of the centrifugal force to the thermal pressure force, 
 is expressed using our parameters as
\begin{eqnarray}
  \beta&=&\frac{\rho(r) v_\phi(r)^2/r}{c_s^2 d\rho(r)/dr} \nonumber \\ 
 &=&\frac{2\Omega_{c\,0}^2H^2}{c_s^2} \nonumber \\
 &=&2\Omega'^2\frac{1+\alpha/2}{F-2\Omega'^2}.
							\label{eq11}
\end{eqnarray}

To initiate the cloud collapse,
 we added a small density perturbation.
The wavelength of the perturbation is taken equal to that of the most
 unstable Jeans mode.
The linear instability of the rotating, magnetized, isothermal cylindrical
 cloud is studied by \citet{mat94}.
They gave a fitting formula for the most unstable wavelength as follows:
\begin{equation}
 \lambda_{\rm max}\left[\frac{c_s}{(4\pi G \rho_s)^{1/2}}\right]^{-1}
  \simeq \frac{2\pi \left(1+\alpha/2+\beta \right)^{1/2}}
        {0.72 \left[\left(1+\alpha/2+\beta\right)^{1/3}-0.6 \right] F^{1/2}},
							\label{eq12}
\end{equation}
using equations (\ref{eq7}), (\ref{eq9}), and (\ref{eq11}).
Therefore, we take this most unstable mode and the initial density
 is assumed equal to
\begin{equation}
  \rho(z,r)=\rho_0(r)\left[1+A\cos\left(\frac{2\pi z}{\lambda_{\rm max}}\right)\right],
							\label{eq13}
\end{equation}
for $-\lambda_{\rm max}/2 \le z \le \lambda_{\rm max}/2$.
The amplitude of the perturbation $A$ is taken equal to 0.1.
Hereafter, we omit the primes which indicate the non-dimensional quantities,
 unless the quantities are confused with dimensional ones.

From calculations of one-dimensional, non-rotating, non-magnetized cloud,
 the effective equation of state of the gas forming a star is summarized
 as follows \citep{toh82}.
Gas in the molecular cloud core with $\rho\sim 10^4{\rm H_2\ cm}^{-3}$ obeys
 the isothermal equation of state with the temperature of $T_0\sim 10$K.
However, the molecular gas becomes optically thick against the thermal
 emission from the dust, which cools the molecular core,
 after the contraction proceeds and the central density reaches 
 $\rho_c = \rho_{\rm A} \sim 10^{10}{\rm H_2\ cm}^{-3}$ \citep{mas99}.
The gas becomes adiabatic beyond the density and the first core forms
 \citep{lar69}.
Hydrogen molecules (H$_2$) dissociate into atomic hydrogen (H) when its
 temperature reaches $T_{\rm dis}\simeq 10^3$K.
Typical density at which the temperature reaches $T_{\rm dis}$ 
 is equal to  $\rho_{\rm B}= \rho_{\rm A}(T_{\rm dis}/T_0)^{1/(\gamma-1)}\simeq
 10^{15} {\rm H_2\ cm}^{-3}(T_{\rm dis}/10^3{\rm K})^{5/2}
 (T_0/10{\rm K})^{-5/2}$.     
The dissociation ends at the density  
 $\rho_{\rm C} \simeq 10^{19} {\rm H_2\ cm}^{-3}$. 
Finally forms an atomic hydrogen core, which is call the second core by
 \citet{lar69}.
To include these changes, we adopt the multiple polytropes
\begin{equation}
 p=\left\{\begin{array}{ll}
          c_s^2 \rho,
           & \rho < \rho_{\rm A},\\
          c_s^2 \rho_A\left(\frac{\rho}{\rho_{\rm A}}\right)^{7/5}, 
           & \rho_{\rm A}< \rho < \rho_{\rm B},\\
          c_s^2 \rho_A\left(\frac{\rho_{\rm B}}{\rho_{\rm A}}\right)^{7/5}
          \left(\frac{\rho}{\rho_{\rm B}}\right)^{1.1}, 
           & \rho_{\rm B}< \rho < \rho_{\rm C},\\
          c_s^2 \rho_A\left(\frac{\rho_{\rm B}}{\rho_{\rm A}}\right)^{7/5}
          \left(\frac{\rho_{\rm C}}{\rho_{\rm B}}\right)^{1.1}
          \left(\frac{\rho}{\rho_{\rm C}}\right)^{5/3},  & \rho > \rho_{\rm C},
\end{array} \right.
\label{eqn:toh82}
\end{equation}  
with $\rho_{\rm A}=10^{10}{\rm H_2\,cm^{-3}}$,
 $\rho_{\rm B}=10^{15}{\rm H_2\,cm^{-3}}$,
 and $\rho_{\rm C}=10^{19}{\rm H_2\,cm^{-3}}$.

The basic equations to be solved are the magnetohydrodynamical equations
 and the Poisson equation for the gravitational potential.
In cylindrical coordinates ($z$, $r$, $\phi$)
 with $\partial / \partial \phi=0$, the equations are expressed as follows:
\begin{equation}
 \frac{\partial \rho}{\partial t}+\frac{\partial}{\partial z}(\rho v_z)+\frac{1}{r}\frac{\partial}{\partial r}(r\rho v_r)=0,
                                                                  \label{eq14}
\end{equation}
\begin{eqnarray}
   \frac{\partial \rho v_z}{\partial t}&+&\frac{\partial}{\partial z}( \rho v_z v_z) +
  \frac{1}{r}\frac{\partial}{\partial r}(r\rho v_z v_r)= \nonumber \\ 
  &- & c_s^2 \frac{\partial \rho}{\partial z}
     -\rho\frac{\partial \psi}{\partial z}
     +\frac{1}{4\pi}\left[-\frac{\partial B_\phi}{\partial z}B_\phi
                         -\left(\frac{\partial B_r}{\partial z}
                         -\frac{\partial B_z}{\partial r}\right)B_r\right],
                                                                  \label{eq15}
\end{eqnarray}
\begin{eqnarray}
 \frac{\partial \rho v_r}{\partial t}&+&\frac{\partial}{\partial z}( \rho v_r v_z) +
  \frac{1}{r}\frac{\partial }{\partial r}(r\rho v_r v_r)= \nonumber \\ 
  &- & c_s^2 \frac{\partial \rho}{\partial r}
     -\rho\frac{\partial \psi}{\partial r}
     +\frac{1}{4\pi}\left[-\frac{1}{r}\frac{\partial}{\partial r}(r B_\phi)B_\phi
                         +\left(\frac{\partial B_r}{\partial z}
                               -\frac{\partial B_z}{\partial r}\right)B_z\right], 
                                                                  \label{eq16}
\end{eqnarray}
\begin{equation}
  \frac{\partial \rho r v_\phi}{\partial t}+
   \frac{\partial}{\partial z}( \rho r v_\phi v_z) +
   \frac{1}{r}\frac{\partial }{\partial r}(r \rho r v_\phi v_r) 
  =\frac{r}{4\pi}\left[
      \frac{1}{r}\frac{\partial}{\partial r}(rB_\phi)B_r
      +\frac{\partial B_\phi}{\partial z}B_z   \right] 
                                                                  \label{eq17}
\end{equation}

\begin{equation}
  \frac{\partial B_z}{\partial t}=\frac{1}{r}\frac{\partial }{\partial r}[r(v_zB_r-v_rB_z)],
                                                                  \label{eq18}
\end{equation}
\begin{equation}
  \frac{\partial B_r}{\partial t}=-\frac{\partial }{\partial z}(v_zB_r-v_rB_z),
                                                                  \label{eq19}
\end{equation}
\begin{equation}
  \frac{\partial B_\phi}{\partial t}=
     \frac{\partial }{\partial z}(v_\phi B_z - v_z B_\phi)
     -\frac{\partial }{\partial r}(v_r B_\phi -v_\phi B_r),
                                                                  \label{eq20}
\end{equation}
\begin{equation}
  \frac{\partial^2 \psi}{\partial z^2}
 + \frac{1}{r}\frac{\partial }{\partial r}\left(r\frac{\partial \psi}{\partial r}\right) =
   4 \pi G \rho,
  \label{eq21}
\end{equation}
 where the variables have their ordinary meanings.
Equation (\ref{eq14}) is the continuity equation;
equations (\ref{eq15}), (\ref{eq16}) and (\ref{eq17}) are the
 equations of motion.
The induction equations for the poloidal magnetic fields are
 equations (\ref{eq18}) and (\ref{eq19}) and for the toroidal magnetic field
 is equations (\ref{eq20}).
The last equation (\ref{eq21}) is the Poisson equation.

The MHD equations were solved using \citet{van77}'s monotonic interpolation
 and the constrained transport method by \citet{eva88} with the method
 of characteristics (MOC) modified by \citet{sto92}.
To ensure the specific angular momentum $\rho r v_\phi$ and
 the toroidal magnetic fields $B_\phi$ are convected similar to the density,
 advections of such values are calculated using 
 the consistent advection which is first pointed out by \citet{nor80}.
We rewrote equations (\ref{eq17}) and (\ref{eq20}) as
\begin{mathletters} 
\begin{eqnarray}
 \frac{\partial B_\phi}{\partial t}
  &=&-\frac{\partial}{\partial z}
     \left[\left(\frac{B_\phi}{\rho}\right) \left(\rho v_z\right)\right]
-\frac{\partial}{\partial r}
     \left[\left(\frac{B_\phi}{\rho}\right) \left(\rho v_r\right)\right],\\
 \frac{\partial \rho r v_\phi}{\partial t}
  &=&-\frac{\partial}{\partial z}
     \left[\left(r v_\phi \right) \left(\rho v_z\right)\right]
-\frac{\partial}{\partial r}
     \left[\left(r v_\phi \right) \left(\rho v_r\right)\right].
\end{eqnarray}
\end{mathletters}
To evaluate the right-hand-side of equations (\ref{eq17}) and (\ref{eq20}),
 we chose the same numerical mass flux $\rho${\bf v} used in 
 equations (\ref{eq14}) and (\ref{eq15})
 and multiplied $B_\phi/\rho$ and $r v_\phi$ to the mass flux
 to get numerical flux for the angular momentum density and
 the toroidal magnetic fields. 
To solve the Poisson equation (\ref{eq21}), we adopted MILUCGS 
[modified incomplete LU decomposition preconditioned
 conjugate gradient squared method: \citet{mei77}; \citet{gus78}].

The gravitational contraction proceeds in a significantly
 non-homologous way.
Thus, to see the late phase evolution, 
 high spatial resolution is needed especially near the center.
Similarly to the previous papers \citep{tom96,tom98,tom00},
 we adopt the nested grid method \citep{ber84,ber89}.
In this method, a number of levels of grids with different spacings
 are prepared; finer grids cover the central high-density portion
 and the coarser ones cover the cloud as a whole. 
The grids are named as L0 (the coarsest), L1, L2, \ldots and
 the grid spacing of L$n$ is chosen equal to a half of that of L$n-1$,
 that is, the grid spacings of L$n$ is equal to 
 $\Delta z_n=\Delta z_0 2^{-n}$ and $\Delta r_n=\Delta r_0 2^{-n}$.
The L0 grid covers the region
 $-\lambda_{\rm max}/2 \le z \le \lambda_{\rm max}/2$
 and $0 \le r \le \lambda_{\rm max}$.

In the nested grid method, the boundary condition is applied to the outer boundary of the
 coarsest grid L0, on which we adopted the fixed boundary condition at $r=\lambda_{\rm max}$
 and the periodic boundary condition at $z \pm \lambda_{\rm max}/2$.
The code has been tested comparing the results obtained with and without
 the nested grid technique [for detail see \citet{tom96}].
The the most unstable growth rates of perturbations are
 slightly different ($\simeq 8\%$):
 $\rho_c=10^4$ was attained at $t=1.25$ for without nested grid (but
 using relatively large number of grid points 400$\times$400), but
 at $t=1.351$ in the calculation with the nested grid method.
This is mainly due to the fact the eigen-mode of the most unstable mode
 is as wide as the whole size of the numerical box and 
 this mode is captured better
 by a simple calculation with a large number of zones (400$\times$400)
 rather than the L0 grid (64 $\times$ 64) of the present scheme.
However, after the perturbation has grown to be non-linear, the growth rate
 agrees well.

\section{Results}
\label{sec3}

\subsection{Dynamical Contraction}
\label{sec3.1}

In Model A, we calculated the evolution with $\alpha=1$, $\Omega_0=5$, $F=100$,
 and $\rho_s=10^2{\rm H_2\,cm^{-3}}$. 
We summarized the adopted parameters in Table \ref{table2}.
Similar to the {\it non-rotational} magnetized cloud [see Figs. 2b and 2c of \cite{tom96}],
 the cylindrical cloud fragments into prolate spheroidal shape 
 whose wavelength is equal to $\lambda_{\rm max}$.
This prolate spheroidal shape coincides with the structure expected
 from the linear stability analysis by \citet{mat94}. 
Next, this density enhanced region begins to contract along the major axis of the cylindrical cloud,
 since the magnetic fields are assumed to run parallelly to the major axis.
Finally it forms a contracting disk (pseudo-disk) perpendicular to the magnetic field
 lines [Fig. 2d of \cite{tom96}].
The snapshot at this stage is shown in Figure \ref{fig1}.
Using the conversion factor shown in Table\ref{table1},
the epoch $t=0.6066\tau_{\rm ff}$ corresponds to
 $1.06 {\rm Myr}(\rho_s/10^2{\rm H_2\ cm^{-3}})^{-1/2}$ from the beginning of calculation.
Respective panels of Figure \ref{fig1} have different spatial coverage.
Figure \ref{fig1}a, which shows L1, captures global structure of contracting disk
 extending horizontally which is perpendicular to the magnetic field lines. 
The spatial resolution of L1 is so limited that there seems no internal 
 structures in the contracting disk.
However, L5 which has 16 times finer resolution than L1 shows
 shock fronts facing outer directions extend parallel to the $r$-direction.
Fronts near $z\simeq \pm 0.02 H$  are the fast-mode
 MHD shock front, because the magnetic fields bend toward the front
 passing the shock front.   
We can see another density jump near $z\sim 0.01 H$ (hereafter we will omit
 the sign $\pm$ and mention
 the upper half part of the figure since the structure is symmetric).

The shock fronts parallel to the disk are known in non-magnetized 
 rotating isothermal clouds \citep{nor80,mat97}.
This is not due to the rotation; multiple shock fronts are also found
 in the contracting magnetized cloud without rotation \citep{nak99}.
However, situation becomes a bit complicated in this cloud. 
In the outer region $z \gtrsim  0.02 H$, the magnetic field lines run almost
 vertically ($B_z \gg B_r$ and $B_\phi$).
Passing the MHD fast shock, in the intermediate region 
 ($0.01 H \lesssim z \lesssim  0.02 H$),
the radial and toroidal components are amplified
 and the density increases compared to the outer region.  
Finally after passing another front near $z \simeq  0.01 H$,
 the toroidal component $B_\phi$ decreases.
That is, since the magnetic field lines deflect
 departing from the front at the second front, this is the slow MHD shock.
The density range captured in this panel is from
 $10^{2.3}\rho_s=10^{4.3}(\rho_s/100{\rm H_2\ cm}^{-3})$
 to $10^{7.3}\rho_s=10^{9.3}(\rho_s/100{\rm H_2\ cm}^{-3})$.
It should be noticed that these shocks occurs in the isothermal gas.
This phase is called ``run-away collapse,'' in which the central
 density ($\rho_c$) increases greatly in a finite time-scale.

Figure \ref{fig1}c shows the structure captured by L10.
Almost all the gas in this figure is isothermal.
However, a central small part of the contracting disk
 $r \lesssim 6 \times 10^{-4}H$ and $|z|\lesssim 1 \times 10^{-4}H$
 enters the adiabatic regime.
At this stage, we can see another density jump is forming just outside
 the adiabatic part of the disk.
It seems to grow into an accretion shock front, since
 it is known that an accretion shock forms outside the core
 when the adiabatic core develops \citep{lar69}.
This is easily understood as follows: the adiabatic gas with 
 specific heat ratio $\gamma > 4/3$ has a hydrostatic equilibrium
 irrespective of its mass.
The scale-height in the $z$-direction of the core becomes larger than 
 that of the disk.
The adiabatic part of the contracting disk forms a spherical static core.

Figure \ref{fig2} shows the cross-cut view along two axes (panel a: along
 the disk mid-plane $z=0$ and panel b: along the $z$-axis $r=0$). 
The lines with 0.6066 represent the stage shown in Figure \ref{fig1}.   
At this stage ($t=0.6066\tau_{\rm ff}$),
 inflowing gas is almost isothermal ($\rho < \rho_{\rm A}$).
In Figure \ref{fig2}a, the radial distributions of the density,
 the magnetic flux density, and the radial and toroidal components of velocity
 are shown.
We can see that density and magnetic flux density distributions are
 approximately expressed by power-laws 
 as $\rho\propto r^{-2}$ and $B_z \propto r^{-1}$ except for the central part.
At $t=0.6067\tau_{\rm ff}$, the density in the core exceeds
 $10^9\rho_s=10^{11}(\rho_s/100{\rm H_2\,cm^{-3}}) {\rm H_2\,cm^{-3}}$.
And at $t=0.6068\tau_{\rm ff}$,
 it reaches $10^{10}\rho_s=10^{12}(\rho_s/100{\rm H_2\,cm^{-3}}){\rm H_2\,cm^{-3}}$.
At this stage, a radially outward-facing shock front is seen even
 inside the disk; infall motion is abruptly decelerated
 and the density and magnetic flux density are amplified. 
This shows that a compact core is formed inside the accretion shock front.
The central density increases with time and inside 
 $r \lesssim 1.8\times 10^{-4}H$ adiabatic gas ($\rho > \rho_{\rm A}$)
 distributes.
The size of the core is equal to $r \sim 1.9\times 10^{14}(c_s/190 {\rm m\,s^{-1}})
(\rho_s 10^2{\rm H_2\,cm^{-3}})^{-1/2}$.
This reduces with time since the mass of the core increases by the effect of continuous accretion.

Before the shock front is formed $t < 0.6067 \tau_{\rm ff}$,
 the radial inflow velocity takes the maximum about $\simeq 2.5 c_s$
 near $r\simeq 7\times 10^{-3}H$.
For the Larson-Penston self-similar solution 
 for the spherically symmetric dynamical collapse \citep{lar69,pen69},
 this maximum inflow speed is expected equal to $\simeq 3.28c_s$.
On the other hand, it equals to $\simeq 1.736 c_s$ for non-rotating isothermal
 {\em disk} \citep{sai98}.
Therefore, it is shown that the actual inflow speed ranges 
 between the spherically symmetric self-similar solution
 and the axially symmetric thin disk solution.    
After the shock front is formed around the core,
 the inflow velocity takes the maximum
 just outside the shock front and the maximum speed increases with time.
Inflow motion is accelerated toward the shock front.
Similar acceleration is also seen in the toroidal velocity, $v_\phi$.
Before the core formation the toroidal speed $v_\phi$ takes the maximum near 
 $r\simeq 2.5 \times 10^{-3}H$.
However, $v_\phi$ increases toward the accretion shock after core formation.
At $t=0.6068\tau_{\rm ff}$ it reaches $v_\phi\simeq 3 c_s$ (see also Fig.1 of \citet{tom98}).
  
Structure seen in the cross-cut along the $z$-axis is more complicated
 (Fig.\ref{fig2}b). 
Two shock fronts mentioned earlier (Fig.\ref{fig1}b) correspond to
 the jumps near $|z| \simeq 0.02 H$ and $\simeq 0.005 H$\footnote{The shape
 of the inner slow MHD shock is concave.
 On the $z$-axis it is found near $z\simeq 0.005 H$, while
 departing from the $z$-axis ($r \gtrsim 0.02H$) it is found near 
 $z\simeq 0.01H$.}  
At $t=0.6066\tau_{\rm ff}$, the density and the inflowing velocity
 distributions have no discontinuities besides these two shock fronts.
However, at $t=0.6067\tau_{\rm ff}$, the inflowing velocity distribution
 begins to indicate a clear discontinuity near
 $|z| \simeq 1.5 \times 10^{-4} H$.
This is a newly formed shock front and propagate spatially.
Comparing two curves of $t=0.6067\tau_{\rm ff}$ and $t=0.6068\tau_{\rm ff}$,
 it is shown that this shock front breaks into two fronts and
 the inner one ($|z| \simeq 1.5 \times 10^{-4} H$) is standing still,
 while another outer one ($|z|\simeq 5 \times 10^{-4}H$) is
 propagating outwardly.
These two shock fronts are outwardly facing.
Thus the inwardly propagation of the inner fronts is due to the infalling
 gas motion.

\subsection{Outflow}

Figure \ref{fig3} illustrates the structure at $t=0.6069\tau_{\rm ff}$.
Although the gas is inflowing both inside and outside of the disk at 
 $t=0.6067\tau_{\rm ff}$ (Fig.\ref{fig2}),
 at this stage $t=0.6069\tau_{\rm ff}$ ($\tau=3.2\times 10^{-4}\tau_{\rm ff}$)\footnote{$t$ represents
 the time from the beginning of calculation but $\tau$ represents the time after the core formation.
 We assumed that the core consists of the gas with density $\rho > \rho_{\rm A}$.}
 prominent outflow is formed.
It is shown that the flow pattern is completely changed in
 $\Delta t\simeq 2\times 10^{-4}\tau_{\rm ff} \sim 400$yr.
Outflow sweeps a sphere with radius of $r \lesssim 1.2 \times 10^{-3}H$ (Fig.\ref{fig3}a).
Figure \ref{fig3}b indicates that
 the gas near the disk surface flows inwardly for $r \gtrsim 2\times 10^{-4}H$.
However, the direction of the flow is changed upwardly near $r \simeq 2\times 10^{-4}H$.
Finally this gas is ejected.
While the gas near the mid-plane of the disk ($|z|\lesssim 1\times 10^{-5}H$) continues to contract.
This is reasonable because the total amount of angular momentum
 in one magnetic flux tube must be conserved in the axisymmetric
 ideal MHD simulation;
for the outflow gas to get angular momentum, a part of the 
 gas in the same magnetic flux tube has to lose its angular momentum
 and falls further.
In the acceleration process of the gas, the angular momentum is transferred
 from the gas near the mid-plane to the gas near the surface of the disk.
Considering the angular rotation speed, the angular momentum is transferred from the fast-rotating
 mid-plane to the slowly rotating surface gas.

From Figures \ref{fig1}c and \ref{fig3}a (both show the structure captured by L10),
 we can see the magnetic field lines run completely differently comparing
 before ($\rho_c < \rho_{\rm A}$: Fig.\ref{fig1}c) and after ($\rho_c > \rho_{\rm A}$: Fig.\ref{fig3}a)
 the adiabatic core formation.  
That is, in the isothermal runaway collapse phase (Fig.\ref{fig1}c)
 the magnetic fields lines run vertically, in other words,
 perpendicularly to the disk.  
In contrast, after the adiabatic core is formed, 
 the disk continues to contract and drags the magnetic field lines inwardly. 
Thus the angle between the magnetic field lines and the disk decreases.

Figure \ref{fig3}b is a close-up view whose spatial resolution is 4-times finer than that of 
 Figure \ref{fig3}a.
This panel shows us that the angle between the flow and the disk
 is about $\simeq 45\deg$.
The reason why the outflow begins after the core formation is 
 related to the angle between the magnetic field lines and the
 disk, $\theta_{\rm mag}$.
\citet{bla82} have pointed out that for a cold gas rotating with the Keplerian speed 
 to get angular momentum
 from the Keplerian disk via infinitely strong magnetic fields, $\theta_{\rm mag}$
 must be smaller than a critical value $\theta_{\rm cr} = 60 \deg$.
This is understood as follows: 
Consider the gas on one magnetic flux tube.
When the magnetic flux tube rising steeply from the disk as
 $\theta_{\rm mag} > \theta_{\rm cr}$, 
 the gas has to climb the effective potential well even if it rotates
 with the same angular speed of the Keplerian disk.
On the other hand,  when $\theta_{\rm mag} < \theta_{\rm cr}$,
 gas can escape from the gravitational well by getting angular momentum
 from the disk, if the gas has the same angular speed of the Keplerian disk.
Although the exact value of $\theta_{\rm cr}$ 
 depends on the disk rotation speed and the disk-to-central star mass ratio,  
 small angle is preferable to acceleration.
It should be noted that
 this configuration is achieved only after the core formation \citep{tom98}.
 
Figure \ref{fig3}c shows the ratio of the toroidal magnetic component
 to poloidal one by contour lines, which is overlaid on the magnetic field lines.
Since the toroidal-to-poloidal ratio is as small as $\sim 0.6$ in the disk,
 the disk is poloidal-dominated.
However, in the region where the gas flows outwardly the toroidal
 component grows and the toroidal-to-poloidal ratio reaches $\gtrsim 5-8$.
The outflow gas is toroidal-field-dominated.
The coincidence of acceleration region with the toroidal-dominant region
 seems to indicate that the toroidal fields play an important role
 to accelerate the gas.
The toroidal component of the Lorentz force, 
\begin{eqnarray}
F_\phi&=&\frac{1}{c}\left(j_zB_r-j_rB_z\right)\nonumber \\
      &=&\frac{1}{4\pi}\left(\frac{1}{r}\frac{\partial rB_\phi}{\partial r}B_r
                   +\frac{\partial B_\phi}{\partial z}B_z\right),
\end{eqnarray}
works below this toroidal-dominant region, that is,
 $z \lesssim 5\times 10^{-5}H$. 
This toroidal component $F_\phi$ accelerates the toroidal velocity $v_\phi$
 and resultant toroidal motion amplifies the toroidal component of the magnetic fields.
Outflow speed exceeds the sound speed and the fastest
 speed reaches $v_{\rm out}\simeq 7.5 c_s$ at this time.     
It increases with time.

\subsection{Effect of the Hardness of the Polytropic Gas}

Although the outflow seems to continue, the further evolution is hard to
 study, because the time-scale (the free-fall time-scale at the central core)
 becomes shorter and shorter.
Therefore, we study model AH with a constant polytropic 
 index larger than that of model A. 
In models AH1 (Fig.\ref{fig4}a) and AH2 (Fig.\ref{fig4}b),
 the polytropic indices are chosen 
 $\Gamma=2$ and $\Gamma=5/3$, respectively, for $\rho > \rho_{\rm A}$.
(Models whose name have ``H'' have simple polytropic relation
 with $\Gamma=2$ or $\Gamma=5/3$ for $\rho > \rho_{\rm A}$.)   
Due to the hard polytropic index,
 the size of the adiabatic core, whose surface  is determined by the jump in $v_r$,
  becomes large;
  for example at $t=0.6069\tau_{\rm ff}$ the size is equal to
  $r_{\rm c}\simeq 4\times 10^{-4} H$ for model AH1
  and $r_{\rm c}\simeq 3\times 10^{-4} H$ for model AH2
  while it is $r_{\rm c}\simeq 1.5\times 10^{-4}H$ for model A.
Similar to model A,
 just outside the core, outflow begins to be accelerated.
The region which the outflow sweeps expands and the surface
 which separates inflow and outflow forms another MHD shock front.    
And the expansion of the front is very similar to that of model A
 (the front reaches $z\simeq 1 \times 10^{-4}H$ at this time which is 
 similar to model A).

To see the similarity in more detail,
 we calculated the mass of the core 
\begin{equation}
  M_{\rm core}\equiv \int_{\rho > \rho_{\rm A}}\rho dV,
\end{equation}
for models A, AH1, and AH2.
These are equal to $0.1407c_s^3/(4\pi G)^{3/2}\rho_s^{1/2}$,
 $0.1504c_s^3/(4\pi G)^{3/2}\rho_s^{1/2}$, and $0.1450c_s^3/(4\pi G)^{3/2}\rho_s^{1/2}$
 at the time $t=0.6069\tau_{\rm ff}$ ($\tau=3.2\times 10^{-4}\tau_{\rm ff}$)
 for  models A, AH1 and AH2, respectively.
At later epoch $t=0.6085\tau_{\rm ff}$ ($\tau=2\times 10^{-3}\tau_{\rm ff}\simeq 3200 
 (\rho_s/10^2{\rm H_2\,cm^{-3}})^{-1/2}$yr),
 $M_{\rm core}\simeq 0.4697c_s^3/(4\pi G)^{3/2}\rho_s^{1/2}
  \simeq 0.11 (c_s/190{\rm m\,s}^{-1})^3(\rho_s/10^2 {\rm H_2\,cm^{-3}})^{-1/2}M_\odot$
 (model AH1) and $M_{\rm core}\simeq 0.4104c_s^3/(4\pi G)^{3/2}\rho_s^{1/2}
 \simeq 0.09 (c_s/190{\rm m\,s}^{-1})^3(\rho_s/10^2 {\rm H_2\,cm^{-3}})^{-1/2}M_\odot$ (model AH2).
From these results, it is shown that 
 the core mass increases with time due to the continuous accretion and 
 the mass does not depend on the exact equation of state in the core.
This is understood as follows: the core mass is determined
 by the accretion rate of the {\em isothermal gas} which is independent from
 the polytropic $\Gamma$ in the core.

The gravity by the core has an effect on the outer inflow and outflow.
Since the effect depends only on its mass,
 the difference of the polytropic index of the core does not play an 
 important role for the inflow and outflow.    
Therefore, we will study this model AH to see the long time evolution
 of the outflow.
 
In Figure \ref{fig4}c, the snapshot at 
 $t=0.6105\tau_{\rm ff}$ ($\tau=4\times 10^{-3} \tau_{\rm ff}$) is plotted for model AH1.
Comparing this with Figure \ref{fig1}b for model A (both have the same resolution
 but for different epochs),
 it is shown that the shock front which separates the inflow and the outflow
 passed the slow-mode MHD shock near
 $z\simeq 0.01 H$ and has just reached the outer fast-mode shock front
 near $z\simeq 0.02 H$.
The evolution of model AH2 is essentially the same.
The maximum speed of the outflow reaches
  $\sim 8 c_s\sim 2{\rm km\,s^{-1}}(c_s/0.2 {\rm km\,s^{-1}})$.
This maximum speed seems smaller than that observed in the molecular outflow.
Since the mass accumulated in the core is only equal to
 $\sim 0.1(c_s/190{\rm m\,s^{-1}})^3(\rho_s/100{\rm H_2\,cm^{-3}})^{-1/2}M_\odot$,
 the outflow speed seems to be much faster than this value, when the mass has grown to typical T Tauri stars.
Figure \ref{fig3}b indicates the outflow is accelerated near the core 
 and the opening angle of the outflow in this region is wide.
However, departing from the acceleration region
 the flow changes its direction toward the $z$-axis.
Figure \ref{fig4}c show that the opening angle decreases as the outflow proceeds.  
This indicates the flow is collimated.

\subsection{Effect of the Initial Rotation Speed}

To see the effect of the initial rotation speed,
 we compare models AH1 ($\Omega_0=5\tau_{\rm ff}^{-1}$),
 BH ($\Omega_0=1\tau_{\rm ff}^{-1}$), 
 and CH ($\Omega_0=0.2\tau_{\rm ff}^{-1}$).
These models have the same magnetic field strength, $\alpha$.
In Figure \ref{fig5} panels (a-c), the structures at the final epoch of the isothermal
 run-away collapse phase are plotted for respective models.
Models BH and CH indicate no prominent discontinuity in L6 
 (Figs. \ref{fig5}b and c), while model AH1 has several shock fronts 
 as described in $\S$\ref{sec3.1}.
However, in L10 (not shown), there are discontinuities near $z\simeq 6 \times 10^{-4}H$
 (model BH) and  $z\simeq 5 \times 10^{-4}H$ (model CH) as well as in model AH1
 ($z\simeq 1\times 10^{-4}H$).
Comparing panels (b) and (c), distributions of the 
 density and magnetic field lines are similar each other.
This indicates that the evolution in the isothermal phase is
 slightly dependent on the initial angular momentum if $\Omega_0 \lesssim 1\times 
 \tau_{\rm ff}^{-1}$. 

Figures \ref{fig5}d-f show the structure at the age $\tau=4.5\times 10^{-3}
 \tau_{\rm ff}$ after the core formation.
In all models the outflows are formed.
However, the size of the region swept by the outflow is different
 for each model.
With increasing $\Omega_0$, more energetic outflow is driven.
From the flow vectors,
 it is shown that model AH1 (Fig.\ref{fig6}d) forms a bit more collimated outflow
 than models BH and CH  (Fig.\ref{fig6}e and f).
This seems to correspond to the differences in density distribution and
 magnetic field configuration.
That is, in model AH1 (also A and AH2) there is a relatively thick disk seen in L6 which
 is bounded by the shock fronts.
This thick disk seems to confine the outflowing gas in model AH1. 
While, in models BH and CH the disk is relatively thin, which seems to
 make the flow isotropic.
Further, the opening angle of the magnetic field lines
 in models BH and CH is larger than that of model AH1.
This causes the flow also open.

Difference between models BH and CH comes from the fact that the epochs when 
 the outflow begins are different.
Since the initial angular momentum in model BH is five-times larger than
 that of model CH, in model BH the outflow begins earlier than model CH.
At $\tau\simeq 6\times 10^{-3}\tau_{\rm ff}$, however,
 even in model CH the top of the outflow reaches $z\simeq 0.02 H$
 and the structure looks very similar to model BH at $\tau \simeq 4.5 \times 10^{-3}\tau_{\rm ff}$ 
 (Fig.\ref{fig5}e).

At the epoch when Figure \ref{fig5}d-f is taken
 ($\tau=4.5\times 10^{-3}\tau_{\rm ff}\simeq 8\times 10^3{\rm yr}[\rho_s/10^2{\rm H_2\,cm^{-3}}]^{-1/2}$),
 the mass in the adiabatic core reaches
 $M_{\rm core}\simeq 0.667 c_s^3/(4\pi G)^{3/2} \rho_s^{1/2}$ (model AH1), 
 $\simeq 1.311 c_s^3/(4\pi G)^{3/2}\rho_s^{1/2}$ (model BH),
 and $\simeq 1.469 c_s^3/(4\pi G)^{3/2}\rho_s^{1/2}$ (model CH).
The instantaneous rate of mass accretion to the adiabatic core for
 each model attains
 $\dot{M}_{\rm acc} \equiv dM_{\rm core}/dt
  \simeq 110 c_s^3/(4\pi G)$ (model AH1),
 $\simeq 180 c_s^3/(4\pi G)$ (model BH), and
 $\simeq 220 c_s^3/(4\pi G)$ (model CH), respectively.
Therefore, the core mass is approximately proportional to the mass accretion
 rate and the mass accretion rate is larger for models with smaller
 angular rotation speed $\Omega_0$.

Accretion rate expected from the inside-out collapse model \citep{shu77} is
 equal to $0.975 c_s^3/G=12.25c_s^3/(4\pi G)$.
Therefore, the accretion rates calculated here are $9-18$ times larger than that expected by the
 inside-out collapse model.

Consider the reason why the mass of the core decreases with increasing
 $\Omega_0$.
Since the gas is supplied to the core mainly through the disk,
 we will consider the mass inflow/outflow transported in the disk. 
At that time, the gas disk can be divided into three regions.
Outermost region is occupied with isothermal gas and the gas is 
 contracting or inflowing (pseudo-disk).
Therefore the inflow mass rate $(\dot{M}_{\rm in})_{\rm outermost} > 0$ and
 the outflow mass rate $(\dot{M}_{\rm out})_{\rm outermost}=0$ in this outermost region.
Inside of this region, outflow is generated, although a large part of the gas
 is still inflowing.
Therefore in this middle region,  
 the inflow rate is smaller than that of the outermost region,
 $(\dot{M}_{\rm in})_{\rm middle}\lesssim (\dot{M}_{\rm in})_{\rm outermost} 
  \simeq (\dot{M}_{\rm in})_{\rm middle} + (\dot{M}_{\rm out})_{\rm middle}$,
 and the excess mass is transported to the outflow,
 $(\dot{M}_{\rm out})_{\rm middle} > 0$. 
Innermost is the adiabatic core.
Since the mass accretion rate to the core is equal to the net mass inflow
 rate from the middle region,
 $\dot{M}_{\rm acc}\simeq (\dot{M}_{\rm in})_{\rm middle}$.

Mass inflow driven by the self-gravity is more important in a model with
 small $\Omega_0$ in which the self-gravity is ineffectively counterbalanced
 with the centrifugal force.
Therefore  $(\dot{M}_{\rm in})_{\rm outermost}$ becomes larger for slow
 rotator.
This is the first effect of the rotation.

Furthermore, the outflow brings away appreciable amount of gas.
As mentioned previously, the outflow is strongly generated in the fast rotator.
Thus, the mass outflow rate increases with increasing $\Omega_0$
 as $(\dot{M}_{\rm out})_{\rm middle} \sim 80 c_s^3/(4\pi G)$ (model AH1),
 $\sim 20 c_s^3/(4\pi G)$ (model BH), and
 $\sim 10 c_s^3/(4\pi G)$ (model CH).
As a result, increasing $\Omega_0$,
 the portion of outflow gas to the inflow gas
 $(\dot{M}_{\rm out})_{\rm middle}/(\dot{M}_{\rm in})_{\rm middle}$
 becomes large as
 $\sim$ 40\% for model AH1,
 $\sim$ 10\% for model BH,
 and $\lesssim$ 5\% for model CH. 
These two effects works cooperatively to reduce the mass accretion rate
 $\dot{M}_{\rm acc} \simeq (\dot{M}_{\rm in})_{\rm middle}$ to the core for the cloud with a
 large $\Omega_0$.

The maximum outflow speed realized in respective figures are equal to 
 $V_{\rm max} \simeq 9.3c_s$ (model AH1), $6.4c_s$ (model BH), and $3.2c_s$ (model CH).
Since the outflow is accelerated by the toroidal magnetic fields which are generated
 by the rotation motion, this $V_{\rm max}$ increases with increasing $\Omega_0$.      

As shown in \citet{tom00}, since the excess angular momentum of the
 inflowing gas is effectively removed by the outflow, the total angular momentum of the first 
 core which is defined as a gas with $\rho > \rho_{\rm A}$ is equal to
 $J_{\rm core}\simeq 9.2\times 10^{-5}c_s H$ contained in the mass of $M=0.67c_s^3/(4\pi G)^{3/2}\rho_s^{1/2}$ (model AH1),
 $\simeq 2.5\times 10^{-6}c_s H$ in $M=1.31c_s^3/(4\pi G)^{3/2}\rho_s^{1/2}$ (model BH),
 and $\simeq 7.4\times 10^{-6}c_s H$ in $M=1.47c_s^3/(4\pi G)^{3/2}\rho_s^{1/2}$ (model CH).
The total angular momenta contained in the first cores are only 1.1\%, 0.07\%, and 0.9\% of
 the initial ones contained in respective mass $M$.

\subsection{Effect of the Magnetic Field Strength}

To see the effect of the magnetic field strength,
 we calculated models NH, DH and EH in which we took
 $\alpha=0$, $0.1$, and $0.01$.

\subsubsection{Model without Magnetic Fields}

Model NH has no magnetic fields.
In Figure \ref{fig6}a,
 a snapshot at $t=0.6977 \tau_{\rm ff}$ captured by L8 is shown
 for model NH.
At this stage, whole the cloud is in isothermal regime
 and the disk experiences the run-away collapse even if the centrifugal
 force may work to support the cloud. 
This confirms the earlier studies by \citet{nor80} and \cite{nar84}.
The physical reason why the centrifugal force does not stop the contraction
 in the isothermal run-away collapse phase
 is explained in \cite{hay80} as follows:
Due to the the centrifugal force, the mass contained in the 
 Jeans scale ($\sim c_s/(G\rho_c)^{1/2}$) from the center is {\em decreasing}
 throughout the collapse.
In this sense the centrifugal force does work!
Only small part of the mass that resides near the center becomes 
 high-density.
But contraction itself continues and the central density rises greatly in a finite time-scale,
 if the isothermal equation of state is valid.
   
Similar to the previous magnetized models,
 a small adiabatic core is formed first.
Since there is no magnetic fields, magnetic braking does not work, however,
 in this model.
Therefore, gas that accretes on the core has relatively large angular momentum
 in contrast to the magnetized model.
As a result, a ring forms by the gas which accreted on the adiabatic core.
Since the specific angular momentum, $j \equiv r v_\phi$, increases further with time,
 radius of the ring grows.   
Another snapshot in panel (b) at $t=0.7011 \tau_{\rm ff}$ 
($\tau\simeq 3.4 \times 10^{-3}\tau_{\rm ff}$) shows the ring
 clearly.
The ring seems unstable for non-axisymmetric perturbation.
This may form a spiral structure similar to that found by \citet{kle99}.
However this is beyond the scope of this paper.
Therefore, it is concluded that 
 a rotating but non-magnetic cloud leads to a rotating ring in the adiabatic accretion stage.

\subsubsection{Models with Magnetic Fields}
\label{sec:weak-B}

To see the effect of the magnetic field strength,
 in Figure \ref{fig7} we compare models BH ($\alpha=1$), DH($\alpha=0.1$),
 and EH ($\alpha=0.01$).
All models have the same initial rotation speed $\Omega_0=1$, and 
 polytropic gamma $\Gamma=2$.
In panels (a)-(c), the structures at the epoch when an adiabatic core
 begins to form are plotted.
Comparing these panels, it is shown that 
 in model B ($\alpha=1$) a flare-up disk is formed whose isodensity lines 
 are departing from the disk mid-plane leaving from the center.
Decreasing the initial magnetic field strength,
 the shape of dense part of the disk becomes rounder.
Similar effect is already reported for non-rotating magnetized cloud collapse
 \citep{tom96}, that is,
 deceasing $\alpha$ the shape of the isothermal contracting disk 
 becomes rounder and finally reaches a sphere for $\alpha=0$. 
  
In panel (d) we plotted a snapshot for model BH at $t=0.7264\tau_{\rm ff}$ 
 ($\tau=4.46\times 10^{-3}\tau_{\rm ff}$) captured in L7.
This is the same as illustrated in Figure \ref{fig5}e but captured in
 L7 which has twice as fine as Figure \ref{fig5}e.  
Figure \ref{fig7}d which shows the structure near the
 root of outflow indicates that it is very similar to that of model AH1.
For example, the outflow leaves from the disk in the direction almost
 parallel to the disk but it changes its direction to the $z$-direction.
Time $t=0.7264\tau_{\rm ff}$ is equivalent to 
 $\tau=4.46\times 10^{-3}\tau_{\rm ff}$.
It is concluded that 
 in a time-scale of $\tau \simeq 4.5\times 10^{-3}\tau_{\rm ff}$,
 the flow pattern is completely changed from the run-away collapse
 to the outflow plus continuous inflow in the disk.  
The outflow gas flows along the surface whose shape resembles a capital letter U.
The flow departs from the disk with a wide opening angle but it change its direction parallel
 to the $z$-axis.

In panel (e), we plotted the structure expected from a model
 with weaker magnetic field (model DH: $\alpha=0.1$ and $\Omega_0=1$).
The snapshot corresponds to the epoch of  $t=0.7836\tau_{\rm ff}$. 
This corresponds to $\tau=4.59\times 10^{-3}\tau_{\rm ff}$ which is similar to the time-scale
 between panels (a) and (d).
In contrast to the previous model BH, 
 the outflow gas forms a sphere and the magnetic field lines
 are folded inside this sphere.
The magnetic field lines are folded by the pinch or hoop stress of the 
 toroidal magnetic field.
The toroidal magnetic field component is confined 
 in the region where the adjacent poloidal magnetic field lines are 
 running in the opposite directions.
For example, the regions around $(r,z) \sim (0.004H, 0.003H)$ and
 $(r,z) \sim (0.006H, 0.003H)$ in Figure \ref{fig7}e.

In this model, the initial poloidal magnetic fields are weak compared
 to model BH.
Therefore, rotational motion amplifies the toroidal fields and their
 strength surpasses easily that of the poloidal ones.
Thus, the hoop stress by the toroidal field pinches efficiently 
 the magnetic field lines.
In the outflow acceleration region, the toroidal component is predominant
 over the poloidal one.
Magnetic field lines are pinched locally and folded.  
As a result, a spherical magnetic bubble is formed  in this process,
 in which the toroidal magnetic field is predominant.

Toroidal component of the magnetic fields is continuously generated
 by the twist motion driven by the disk rotation.
The disk angular momentum is transferred by this process.
As a result, we do not see the ring of model NH which is supported by the centrifugal force.  

For the model with extremely weak field, we calculated model EH ($\alpha=0.01$ and $\Omega=1$).
In panel (f), we plotted the snapshot at 
 $t=0.7830\tau_{\rm ff}$ 
($\tau=4.53\times 10^{-3}\tau_{\rm ff}$).
Density distribution and magnetic field configuration show that 
 the flow in the magnetic bubble is more complicated or turbulent
 in this model.
The shape of the bubble is more elongated than that formed in model DH.
Distribution of toroidal field lines does not show any systematic pattern inside the bubble. 
Size of the bubble both in the $z$- and $r$-directions are smaller than that 
 of models BH and DH.
It is concluded that the size of the outflow region
 increases with increasing the magnetic field strength $\alpha$.
Comparing these three models, it is shown that there are at least two types of outflows.
That is, a laminar U-type flow in which fast moving gas flows along the surface whose shape resembles
 a capital letter U and a turbulent outflow in which the magnetic fields and the velocity are randomly oriented. 

The masses accumulated in $\tau \sim 4.5\times 10^{-3}\tau_{\rm ff}$ are equal to
 $1.3 c_s^3/(4\pi G)^{3/2}\rho_s^{1/2}$ for model BH,
 $0.87 c_s^3/(4\pi G)^{3/2}\rho_s^{1/2}$ for model DH,
 and $0.65 c_s^3/(4\pi G)^{3/2}\rho_s^{1/2}$ for model EH.
This shows that
 the mass accretion rate, $\dot{M}_{\rm acc}$, is an increasing function of
 the initial magnetic field strength, $\alpha$.
This seems strange if we remember $\dot{M}_{\rm acc}$ is an decreasing function of 
 the initial rotation speed, $\Omega_0$,
 since both $\alpha$ and $\Omega_0$ have an effect to counterbalance against the self-gravity. 
This means that the mass inflow rate in the isothermal run-away collapse region,
 $(\dot{M}_{\rm in})_{\rm outermost}$, increases with increasing $\alpha$.
This seems to come from a number of reasons, that is,
 the initial cylindrical cloud becomes massive with increasing $\alpha$. 
Another reason is related to the characteristic wave speed in the magnetized medium.
That is, the characteristic speed of the fast mode MHD wave 
 is equal to $(c_s^2 + B_0^2/4\pi \rho)^{1/2}=c_s(1+\alpha)^{1/2}$ 
 in the case that the wave is propagating perpendicular to the magnetic field lines.
This implies that the mass inflow rates are proportional to $(1+\alpha)^{3/2}$ 
 (A: B: C = 2.83: 1.15: 1.02).
This is not inconsistent with the actual values.

\section{Discussion}

\subsection{Evolution to Form the Second Core}

In the previous section, we introduced a simple polytrope above the critical density
 $\rho > \rho_{\rm A}$ except for model A.
However, this is an approximation to see a long evolution.
In reality, the composite polytrope (eq. \ref{eqn:toh82}) should be used.
Here, further evolution later than that shown in Figures \ref{fig1} and \ref{fig2}
 is shown.
In model R, we used the composite polytrope to include the combined effects such as
 the dynamical compressional heating, the radiative cooling through the dust thermal emissions,
 and the energy loss associated with the H$_2$ dissociation.
This corresponds to the continuation of model A but we consider here a more compact cloud as 
 $\rho_s=10^4{\rm H_2\,cm}^{-3}$.
It should be addressed since $\rho_s$ is assumed hundred-times larger than other models,
 the size scale $H$ is 10-times smaller than other models with $\rho_s=10^4{\rm H_2\,cm^{-3}}$
 as $H\simeq 1.05\times 10^{17}{\rm cm}(c_s/190{\rm m\,s^{-1}})(\rho_s/10^4{\rm H_2\,cm^{-3}})^{-1/2}$.

The evolution up to the first core formation is similar to model A, that is,
 after the equation of state becomes hard, $\rho > \rho_{\rm A}$,   
 collapse of the adiabatic core slows down and isothermal gas begins to accrete on to the core.
In this model since $\rho_s$ is assumed equal to $10^4{\rm H_2\,cm^{-3}}$,
 the size scale $H$ is 10-times smaller than other models with $\rho_s=10^2{\rm H_2\,cm^{-3}}$.
In Figure \ref{fig8}a, we plotted the structure captured by L6 just before the first core formation 
 $t=0.7201\tau_{\rm ff}$.
Since the accretion continues, the mass of the adiabatic core increases with time, which leads
 to collapse the core quasi-statically.
The structure in this stage (at $t=0.7239\tau_{\rm ff}$ or $\tau=3.8\times 10^{-3}\tau_{\rm ff}$)
 is shown in panel (b). 
The radius of the first core shows in this panel is equal to $r\sim 3.5 \times 10^{-4}H\simeq 
 2.45(c_s/190{\rm m\,s^{-1}})(\rho_s/10^4{\rm H_2\,cm^{-3}})^{-1/2}$ AU.
This quasistatic contraction phase ends when the central density reaches $\rho_{\rm B}$.
The structure at this stage ($t=0.724230\tau_{\rm ff}$
 [$\tau=4.132\times 10^{-3}\tau_{\rm ff}$]) is plotted in panel (c).  
Comparing this with panel (b), both of which illustrate L12, it is clear that the adiabatic core
 is contracting.
After that, the second collapse begins.
Since the thermal energy is lost by the process of the dissociation of H$_2$,
 the equation of state is assumed soft ($\gamma \simeq 1.1$) again.
In this phase, the flow is very similar to that realized in the isothermal run-away collapse phase
 ($\rho < \rho_{\rm A}$).
The similarity comes from the fact that the equation of states of the both phases are soft.
In panel (d), we plotted the structure captured in L16 at $t=0.724236\tau_{\rm ff}$ 
 ($\tau=4.138\times 10^{-3}\tau_{\rm ff}$), which represents the typical structure in the second collapse phase.
This continues until the central density exceeds $\rho > \rho_{\rm C}$,
 beyond which another adiabatic core (the second core) is formed shown in panel (e).
The size of the second core is equal to $\simeq 1.3 \times 10^{-6}H\sim 2R_\odot (c_s/190{\rm m\,s^{-1}})
 (\rho_s/10^4{\rm H_2cm^{-3}})^{-1/2}$. 
This meets a similar situation when the first adiabatic core is formed,
 that is, a central part of the gas obeys a harder equation of state and forms a quasistatic core,
 while the outer part obeys a softer equation of state and continues to collapse.
Analogy between the first and the second core leads a bipolar outflow.
In panel (f), it is shown a second outflow is formed around the second core.
This is a snapshot of L16 at $t=0.724237\tau_{\rm ff}$  ($\tau=4.140\times 10^{-3}\tau_{\rm ff}$),
 which resembles the structure seen in Figure \ref{fig3}b.

This model strongly indicates us that there is another kind of outflow accelerated around 
 the second core (the second outflow) besides that formed around the first core (the first outflow).
The outflow speed, $\sim 50c_s$, is much faster than that of the first outflow,
 due to the fast thermal speed in the adiabatic core and
 the deep local gravitational potential. 
Since the time span of the simulation shown here is restricted, we can not trace the evolution further.
However, the simulation predicts that at least two different outflows are formed each of which
 is related to different types of adiabatic cores.
Since the flow speed of the second outflow is much faster than that of the first outflow,
 the respective outflows correspond to the molecular bipolar outflow (the first outflow)
 and the fast neutral wind or optical jets (the second outflow).
The radial size of the outflow is approximately equal to $\sim 2\times 10^{-5}H\sim 2.1\times 10^{12}
 {\rm cm}  (c_s/190{\rm m\,s^{-1}}) (\rho_s/10^4{\rm H_2cm^{-3}})^{-1/2}$ (=0.14 AU or $30 R_\odot$).
This indicates that optical jets are found inside molecular bipolar outflows.  

In model R, we assumed the ideal magnetohydrodynamics.
However, the ionization fraction of high-density gas is quite low and the electric conductivity
 decreases as collapse proceeds.
Calculation of ionization equilibrium \citep{nak86} shows us that after
 $n_{\rm H} \gtrsim 10^{12}{\rm cm}^{-3}$ charged grains are more abundant than ions,
 as a carrier of electric charges.
Since the mass-to-charge ratio of grains is much larger than that of ions, the electric conductivity
 decrease greatly and the magnetic field decays mainly through Joule dissipation.
Therefore, in the late phase of the first core (quasistatic contraction),  
 the magnetic field in the core decreases its strength till the field configuration 
 becomes force-free.
After the dissociation of H$_2$ begins, the temperature of the central region
 is enough high to achieve the thermal ionization of metals.
Thus, in the second collapse phase, coupling of magnetic fields is recovered.
From these, the ideal MHD is consistent in the first and the second collapse phases.
However, the magnetic flux density at the beginning of the second collapse phases
 seems to be much weaker than that obtained here.
As shown in $\S$\ref{sec:weak-B}, when the poloidal magnetic field is weak, the flow pattern is
 different from the case that the magnetic energy is comparable to the thermal one.
Therefore, if the magnetic flux is partly lost from the central part of the first core,
 a turbulent outflow around the second core seems to be formed similar to Figure \ref{fig7}f.        

\subsection{Mass Inflow/Outflow Rate}

Here in this subsection, mass inflow/outflow rate and linear momentum outflow rate are seen more closely.
The outflow mass loss rate through the boundary of each grid level is measured as
\begin{equation}
\dot{M}_{\rm out}({\rm L}n)=\int_{{\rm boundary\ of\ L}n} \rho \max[{\bf v}\cdot {\bf n},0] dS,
\label{eqn:mdotout}
\end{equation} 
where {\bf n} represents the unit vector outwardly normal to the surface.
The integrand means that only the outwardly running mass flux is summed up.
Similarly, the mass inflow rate is calculated as
\begin{equation}
\dot{M}_{\rm in}({\rm L}n)=\int_{{\rm boundary\ of\ L}n} \rho \max[-{\bf v}\cdot {\bf n},0] dS.
\label{eqn:mdotin}
\end{equation} 
These rates are calculated for respective levels of the nested grid system.

The time variations of mass inflow and outflow rates are illustrated for respective models
 in Figure \ref{fig9}.
The time (horizontal axis) is measured from the epoch of the core formation.
Time variations, especially longer time variations, in inflow rates are common for all panels. 
Just after the core formation, the inflowing mass flux decreases with the distance from the center. 
This means that the inner part of the isothermal collapse region is well expressed
 by the \citet{lar69}-\citet{pen69} self-similar solution \citep{ogn99} which leads to
 the mass inflow rate of $47 c_s^3/G$;
Departing from the center, the approximation of the Larson-Penston solution becomes worse
 and the inflow becomes lower.   
At $\tau \sim 4 \times 10^{-3}\tau_{\rm ff}$ the inflow rates for various surfaces
 converge on $\simeq 20 c_s^3/G$, although intense outflow reduces the inflow rate below this value.

We will see each model more closely.
Outflow rate rises in the deeper levels (L$n$ with larger $n$) first and propagates to
 lower levels (L$n$ with smaller $n$).
This indicates the outflow region expands both in $r$- and $z$-directions. 
Comparing models AH1 (panel a), BH (panel b), and CH (panel c), the effects of the initial rotation speed, $\Omega$, are apparent.\\
(1) with increasing $\Omega_0$, the outflow begins earlier.\\
(2) with increasing $\Omega_0$, the mass outflow rate increases; 
 although the ratio $\dot{M}_{\rm out}/\dot{M}_{\rm in}$ is equal to only $\sim 10\%$ in model CH
 ($\Omega_0=0.2$), in model AH1 ($\Omega_0=5$) it attains $\sim 50\%$. 

We compare models BH (panel b), DH (panel d), and EH (panel e) to see the effects of the initial magnetic field strength, $\alpha$.
This shows that with decreasing $\alpha$ (from models BH to DH) the mass outflow rate increases.
However, in model E, in which extremely weak poloidal magnetic fields is assumed, 
 the time variation in the inflow and outflow rates is rather chaotic which is lead by
 the turbulent flow pattern realized in model EH.
Averaging the rates as $<\dot{M}>=\int_0^T\dot{M}dt/T$,
 the rate of model E is $\sim 20-40\%$ {\em smaller} than that of model DH.  
Considering the disk after rotating at a fixed angle, the disk of model EH can generate
 weaker toroidal magnetic fields than model DH, since model EH has only weak poloidal (source)
 magnetic fields. 
Therefore, it is reasonable that the mass outflow rate in model EH, $<\dot{M}_{\rm out}>$, is smaller than      
 that of model DH.

How about the increase from model BH to DH?
This is inconsistent with the above discussion.
This increase seems related to the fact that the flow patterns of models BH and DH are different.
That is, in model DH a magnetic bubble, in which magnetic field lines are folded and amplified, is
 formed and the bubble expands.
In contrast, in model BH, the gas is flowing outward along specific magnetic field lines
 and outflowing gas forms a wall of a chimney whose shape looks like a capital letter U.
Mass outflow rate seems larger in the magnetic bubble type outflow rather than the chimney wall type.

The linear momentum outflow rates are calculated by
\begin{equation}
\left. \frac{d MV}{dt}\right|_{\rm out}({\rm L}n)=\int_{{\rm upper\ and\ lower\ boundaries\ of\ L}n} \rho v_z \max[{\bf v}\cdot {\bf n},0] dS.
\label{eqn:mvdotout}
\end{equation} 
The time variation of momentum outflow rate is shown in Figure \ref{fig10}.

Panels (a), (b), and (c) have a prominent common feature that the momentum outflow rates ($dMV/dt$)
 calculated at the boundaries of L$n$ ($n \le 7$) are much larger than that of the boundaries
 of L$n$ ($n > 7$).
This feature is common in models with $\alpha=1$ (models AH1, AH2, BH, and CH).
However, models with weaker magnetic field does not show this feature.
Recall the fact that models AH1, AH2, BH, and CH
 form a U-shaped outflow and in contrast models DH and EH
 form turbulent magnetic bubble type outflow.
Difference in the momentum outflow rate seems to be related to the outflow pattern.
In the laminar U-shaped outflow,
 the outflow is ejected with a rather wide opening angle and collimated to the symmetric axis.
Since equation (\ref{eqn:mvdotout}) counts the momentum passing the upper and lower boundaries,
 the outflow rate increases after the outflow is collimated and the gas flows parallel to the $z$-axis.
This occurs in the larger scale than L7.

Typical momentum outflow rate for molecular bipolar outflows 
 observed by $^{12}$CO($J=2-1$) \citep{bon96} are distributed between
 $\sim  10^{-4}M_\odot {\rm yr^{-1}\,km\,s^{-1}}$ for objects associated with active Class 0 IR sources
 and $\sim 2\times 10^{-6}M_\odot {\rm yr^{-1}\,km\,s^{-1}}$ for objects associated with late Class 1 IR sources.
This range corresponds to $320 c_s^4/G - 6.5 c_s^4/G$, respectively, if we assume $c_s=190{\rm m\,s}^{-1}$.
Figure 10 shows us that the momentum outflow rate in the range from $\sim 10 c_s^4/G$ to
 $\sim 40 c_s^4/G$ is expected.
Thus, the momentum outflow rate for almost all the CO bipolar outflow sources
 associated with low-mass young stellar objects except for the active early class 0 sources 
 are explained by this model.

Comparing models AH1, BH, and CH, it is shown that the momentum outflow rate increases with $\Omega_0$. 
Comparing mass outflow and momentum outflow rates, models AH1, BH, and CH  indicate that 
 the momentum outflow rate is approximately proportional  to the mass outflow rate,
 which means that the outflow speed is approximately equal irrespective of $\Omega_0$. 
However, changing $\alpha$ is more complicated.
Compare panels (b), (d), and (e).
As shown in Figure \ref{fig9}, model DH ($\alpha=0.1$) shows larger mass outflow rate than model BH ($\alpha=1$).
In contrast, as for the momentum outflow rate, the maximum rate of model BH is larger than that of model DH
 (Fig.\ref{fig10}).
This represents the outflow speed of model BH ($\alpha=1$) is faster than model DH ($\alpha=0.1$).
Panel (e) shows that the momentum outflow rate has a chaotic time variation
 as well as the mass outflow rate (Fig.\ref{fig9}e), for model EH, which seems to correspond
 the turbulent outflow shown in Figure \ref{fig7}f. 

Summarizing the effect of changing $\alpha$,\\
 (1) the mass outflow rate increases with increasing $\alpha\lesssim 0.1$ but decreases after $\alpha \sim 0.1$.\\
 (2) the outflow speed is an increasing function of $\alpha$.\\
 (3) there are distinctly different two types of flow patterns: models with strong magnetic fields
 lead the U-shape outflow, while the flow becomes turbulent for models with weak magnetic fields. 

\subsection{Where Is the Gas Ejected?}

Here, we intend to clear where the outflow is ejected.
In the run-away collapse phase, the rotational motion is relatively unimportant.
After the core formation, rotation motion increases especially around the first core.
This occurs near the centrifugal radius, which is defined for a gas with specific angular momentum $j$ as
\begin{equation}
R_{\rm c} = c_{\rm c} \frac{j^2}{GM},
\label{eqn:eqn-cen-rad}
\end{equation}
where $M$ denotes the mass inside the radius $R_{\rm c}$ and $c_{\rm c}$ represents a numerical factor of the
 order of unity.
Considering the self-similar solution for the run-away collapse phase \citep{sai98},
 the specific angular momentum $j$ is proportional to $M$.
Further a uniform-density cylinder which is rotating with a uniform
 rotation speed $\omega$ has a $j$ distribution proportional to $M$.
Assuming $j=qGM/c_s$ ($q$ is a numerical factor), the centrifugal radius is written 
\begin{equation}
R_{\rm c}=c_{\rm c}\frac{q^2GM}{c_s}.
\label{eqn:eqn-cen-rad-2}
\end{equation}

As shown in section \ref{sec3}, the angular momentum is redistributed in one magnetic flux tube.
That is, gas near the disk surface obtains large amount of the specific angular momentum but
 that near the disk mid-plane loses the angular momentum.   
If the gas near the disk surface has the angular momentum of $j_+ > j$,
 the condition that the effective potential at the centrifugal radius 
\begin{equation}
\Phi(R_{\rm c})=-\frac{GM}{R_{\rm c}}+\frac{j_+^2}{2R_{\rm c}^2}
\end{equation} 
is larger than or equal to zero leads to a minimum angular momentum of
\begin{equation}
j_+ \ge (2 c_{\rm c})^{1/2}j.
\end{equation} 

\section{Summary}

We have explored the evolution of a magnetized interstellar cloud rotating around the symmetric axis.
Following the change in the equation of state of the interstellar gas \citep{toh82},
 the cloud experiences several phases before going to a star, that is,
 the isothermal run-away collapse, the slowly contracting core composed
 of the molecular hydrogen (the first core), the second run-away collapse in the high-density gas
 where the dissociation of hydrogen molecules proceeds, and finally the second core which is
 made up of the atomic hydrogen. 
The magnetized cloud forms a pseudo-disk in these first and second run-away collapses,
 in which a supersonically contracting disk is formed and magnetic field lines are running
 perpendicularly to the disk.
In the pseudo-disk, a number of fast- and slow-mode MHD shock pairs are formed 
 which is extending parallelly to the disk.
Just after the core is formed at the center, an accretion shock front appears which surrounds the core,
 through which the supersonic inflow motion is decelerated.
While the first and second cores are slowly contacting, the outer pseudo disk continues to contract. 
Just outside the accretion shock front, the infall motion is accelerated and thus rotational motion
 becomes important from the conservation of angular momentum. 
By the effect of rotational motion, the toroidal magnetic fields and the poloidal currents are amplified,
 which bring a strong magnetic torque.
The magnetic torque leads the angular momentum transfer from mid-plane to surface of the disk.  
This is the origin of the outflow found in star forming regions.
Large scale bipolar molecular outflows are made outside of the first core, while optical jets and
 fast neutral winds are expected to be accelerated outside of the second core.
Matter lost its excess angular momentum continues to contract further to form a star.

\acknowledgments

This work was supported partially by the Grants-in-Aid (11640231, 10147105) from MEXT 
(the Ministry of Education, Culture, Sports, Science and Technology). 
Numerical calculations were carried out by Fujitsu VPP5000 at the
 Astronomical Data Analysis Center, the National Astronomical Observatory,
 an inter-university research institute of astronomy operated by MEXT, Japan.



\appendix

\section{}

\clearpage

\section*{Table Captions}

\begin{table}[h]
\begin{center}
\caption{Conversion from the non-dimensional to the physical quantities
.\label{table1}}
\begin{tabular}{lll}
Physical Quantities  &  Conversion Factors & Physical Values ($\mu=2.33$)\\
\hline
Velocity  &  $c_s$ &  $190 {\rm m\,s^{-1}}$ \\
Density   &  $\rho_s$  & $10^2{\rm H_2\,cm^{-3}}$ \\
Length    &  $c_s / (4\pi G \rho_s)^{1/2}\equiv H$ &
 $0.341(c_s/190{\rm m\,s}^{-1}) (\rho_s/10^2{\rm H_2\,cm^{-3}})^{-1/2}{\rm pc}$\\
Time      &  $(4\pi G \rho_s)^{-1/2}\equiv \tau_{\rm ff}$ &
 $1.75(\rho_s/10^2{\rm H_2\,cm^{-3}})^{-1/2}{\rm Myr}$\\
Mass      &  $ c_s^3/(4\pi G)^{3/2}\rho_s^{1/2}$ &
 $0.227 (c_s/190{\rm m\,s}^{-1})^3  (\rho_s/10^2{\rm H_2\,cm^{-3}})^{-1/2} M_\odot$ \\
Mass accretion rate & $c_s^3/4\pi G$ &$1.29\times 10^{-7}(c_s/190{\rm m\,s}^{-1})^3M_\odot{\rm yr}^{-1}$\\
  &  $c_s^3/G$\tablenotemark{a} & $1.62\times 10^{-6}(c_s/190{\rm m\,s}^{-1})^3M_\odot{\rm yr}^{-1}$\\
Momentum inflow/outflow rate &$c_s^4/4 \pi G$ &$2.45\times 10^{-8}(c_s/190{\rm m\,s}^{-1})^4M_\odot{\rm yr}^{-1}{\rm km\,s}^{-1}$\\
  &  $c_s^4/G$\tablenotemark{b}   & $3.08\times 10^{-7}(c_s/190{\rm m\,s}^{-1})^4M_\odot{\rm yr}^{-1}{\rm km\,s}^{-1}$\\
Magnetic Field & $c_s \rho_s^{1/2}$ & $3.75(c_s/190{\rm m\,s}^{-1})(\rho_s/10^2{\rm H_2\,cm^{-3}})^{1/2}\mu {\rm G}$\\
\hline
\end{tabular}
\end{center}
\tablenotetext{a}{To meet the conventional normalization, we adopt $c_s^3/G$ in section 4.}
\tablenotetext{b}{To meet the conventional normalization, we adopt $c_s^4/G$ in section 4.}

\end{table}

\begin{table}[h]
\begin{center}
\caption{Model Parameters.\label{table2}}
\begin{tabular}{lccccc}
Model & $\alpha$ & $\Omega_0$ & $\rho_{\rm A}$ & $\rho_s$ & polytrope\\
      &          &            &                &$(\rm H_2\ cm^{-3})$ \\
\tableline
A \ldots & 1 & 5 & $10^8$ & $10^2$ & Realistic \\
AH1 \ldots & 1 & 5 & $10^8$ & $10^2$ & $\Gamma=2$ \\
AH2 \ldots & 1 & 5 & $10^8$ & $10^2$ & $\Gamma=5/3$ \\
B \ldots & 1 & 1 & $10^8$ & $10^2$ &  Realistic \\
BH \ldots & 1 & 1 & $10^8$ & $10^2$ & $\Gamma=2$ \\
CH \ldots & 1 & 0.2 & $10^8$ & $10^2$ & $\Gamma=2$ \\
DH \ldots & 0.1 & 1 & $10^8$ & $10^2$ &  $\Gamma=2$ \\
EH \ldots & 0.01 & 1 & $10^8$ & $10^2$ & $\Gamma=2$ \\
N \ldots & 0 & 5 & $10^8$ & $10^2$ &  Realistic \\
NH \ldots & 0 & 5 & $10^8$ & $10^2$ &  $\Gamma=2$ \\
R \ldots & 1 & 1 & $10^6$ & $10^4$ &  Realistic \\
\tableline
\end{tabular}
\end{center}
\end{table}
\clearpage

\centering
(a)\\
(b)\\
(c)\\
\figcaption[]{Evolution of model A with $\alpha=1$
 and $\Omega_0=5$. 
Snapshots at the time of $t=0.6066\tau_{\rm ff}$ captured by different
 levels are shown: L1 (a), L5 (b), and  L10 (c).
The actual size of the frames of L5 (b) and L10 (c) are,
 respectively, 1/16 and 1/512 smaller than that of L1 (a).  
Magnetic field lines (almost vertical) and isodensity contours are presented
 as well as the velocity vectors by arrows.
Near the right-bottom corner, the logarithm of the maximum and
 the minimum of the densities are numerically shown.
The maximum speed is also shown and the velocity vector corresponding its
 value is illustrated at the lower-left corner by a horizontal arrow.    
\label{fig1}}

\centering
(a)\\
(b)\\
\figcaption[]{Cross-cut views along the equatorial plane (a) and
 the $z$-axis (b).
 The figures show formation of the first core. 
In panel (a), $\log\rho(r,z=0)$ (solid lines), 
 $\log B_z(r,z=0)$ (short dashed lines),
 $v_\phi(r,z=0)$ (long dashed lines),
 and $-v_r(r,z=0)$ (dotted lines) are plotted, while
 in panel (b), $\log\rho(r=0,z)$ (solid lines), 
 $\log B_z(r=0,z)$ (short dashed lines),
 and $-v_z(r=0,z)$ (dotted lines) are plotted.
\label{fig2}
}

\centering
(a)\\
(b)\\
(c)\\
\figcaption{The same as Fig.\ref{fig1} but for the snapshot at
 $t=0.6069\tau_{\rm ff}$.
Panel (a) shows the structure captured by L10 which is the same as
 Fig.\ref{fig1}c.
At this stage, gas begins to outflow from the disk.
Outflow sweeps the sphere $r \lesssim 1.2 \times 10^{-3}H$.
Panel (b) corresponds to L12, which has 4 times finer spatial resolution
 than panel (a).
Panel (c) shows toroidal-to-poloidal magnetic field strength ratio.
Contour lines extending in the direction $\pm 45\deg$ from the disk are for the toroidal-to-poloidal ratios.
Poloidal magnetic field lines are the same as panel (b).
It is shown that the gas outflows along the region
 where toroidal magnetic field is important.\label{fig3}}

\centering
(a)\\
(b)\\
(c)\\
\figcaption{The same as Fig.\ref{fig1} but for models AH1 and AH2.\label{fig4}
Snapshots at the same epoch of Fig.\ref{fig3}a, $t=0.6069\tau_{\rm ff}$, 
 are shown for models AH1 ($\Gamma=2$) and AH2 ($\Gamma=5/3$) in panels (a)
 and (b), respectively.
Although the structure of the core is different,
 the outflow is very similar with each other.
In panel (c), snapshot at $t=0.6105\tau_{\rm ff}$ ($\tau=3.94 \times 10^{-3}\tau_{\rm ff}$
 from the core formation) is plotted for model AH1.
Be careful that the linear size of this panel is 32 times larger than 
panels (a) and (b).  
Comparing with Fig.\ref{fig1}b (the same resolution),
 it is shown that the shock front passed the slow-mode MHD shock and
 has just reached the outer fast-mode shock front.
}

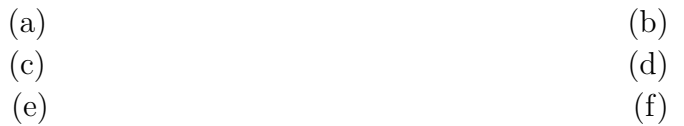
\centering
\hfill\ (a)\hfill\ (b)\hfill \\
\hfill\ (c)\hfill\ (d)\hfill \\
\hfill\ (e)\hfill\ (f)\hfill \\
\figcaption{\label{fig5}
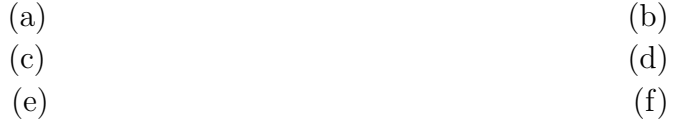
Comparison of models with the same magnetic field strength, $\alpha$,
 but different rotation speed, $\Omega_0$.
Panels (a) and (d) represents the structure of model AH1
 ($\Omega_0=5\tau_{\rm ff}^{-1}$) captured by L6.
Panels (b) and (e) are for model BH.
Model B corresponds to slower rotator $\Omega_0=1\tau_{\rm ff}^{-1}$.
Panels (b) and (e) represents the structure captured by
 L6 at the ages of $t=0.7219 \tau_{\rm ff}$ and $t=0.7264 \tau_{\rm ff}$
 ($\tau=4.5\times 10^{-3}\tau_{\rm ff}$), respectively.
Panels (c) and (f) are for model CH ($\Omega_0=0.2\tau_{\rm ff}$) and show the snapshots at
 $t=0.7262 \tau_{\rm ff}$ and $t=0.7307 \tau_{\rm ff}$
 ($\tau=4.5\times 10^{-3}\tau_{\rm ff}$), respectively.
}


\centering
\hfill\ (a)\hfill\ (b)\hfill \\
\figcaption{The same as Fig.\ref{fig1} but for model NH.\label{fig6}
This model is for a non-magnetized cloud.
In panel (a), 
 a snapshot at $t=0.6977 \tau_{\rm ff}$ captured by L8 is shown.
At this stage, whole the cloud is in isothermal regime.
Another snapshot at $t=0.7011 \tau_{\rm ff}$ is shown in panel (b).
Gas that accretes afterwards forms a ring which is supported
 essentially by the centrifugal force.
}

\centering
\hfill (a)\hfill (b)\hfill \\
\hfill (c)\hfill (d)\hfill \\
\hfill (e)\hfill (f)\hfill \\
\figcaption{
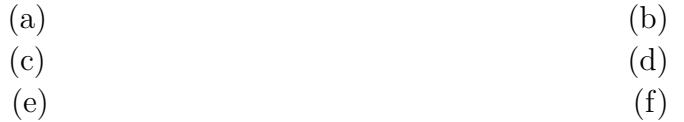
Comparison of models with the same rotation speed $\Omega_0=1$ but
 different magnetic field strength $\alpha$.\label{fig7}
Panels (a), (b), and (c) represent the structure when the adiabatic core 
 begins to form, while panels (d), (e), and (f) represents that after
 a protostar is formed.
In panel (a) we plotted a snapshot for model BH captured by L7
 at $t=0.7219\tau_{\rm ff}$,
 which is the same snapshot shown in Fig.\ref{fig5}b but for different level.
Panel (d) represents the structure of the same model but for the
 protostellar phase, that is,  $t=0.7264\tau_{\rm ff}$ 
 ($\tau=4.46\times 10^{-3}\tau_{\rm ff}$).
This corresponds to Figure \ref{fig5}e.
Panels (b) and (e) illustrate snapshots at $t=0.7790\tau_{\rm ff}$
 and that at $t=0.7836\tau_{\rm ff}$ 
($\tau=4.59\times 10^{-3}\tau_{\rm ff}$), respectively, for model DH
 ($\alpha=0.1$). 
Panels (c) and (f) are for a model with extremely weak magnetic fields 
 (model EH; $\alpha=0.01$).
The snapshots at the epoch $t=0.7784\tau_{\rm ff}$ and $t=0.7830\tau_{\rm ff}$ 
($\tau=4.53\times 10^{-3}\tau_{\rm ff}$) are illustrated in panels (c) 
 and (f), respectively.  
} 

\centering
\hfill (a)\hfill (b)\hfill \\
\hfill (c)\hfill (d)\hfill \\
\hfill (e)\hfill (f)\hfill \\
\figcaption{\label{fig8}
The evolution of model R.
In this model since $\rho_s$ is assumed equal to $10^4{\rm H_2\,cm^{-3}}$,
 the size scale $H$ is 10-times smaller than other models with $\rho_s=10^2{\rm H_2\,cm^{-3}}$.
In panel (a), we plotted the structure captured by L6 just before the first core formation 
 $t=0.7201\tau_{\rm ff}$.
At $t=0.7239\tau_{\rm ff}$ ($\tau=3.8\times 10^{-3}\tau_{\rm ff}$),
 the first core gradually contracts by the effect of continuous mass accretion (panel b).
Finally, $t=0.724230\tau_{\rm ff}$ ($\tau=4.132\times 10^{-3}\tau_{\rm ff}$),
 the central density reaches $\rho_{\rm B}$ (panel c).
After that, the second collapse begins.
In this phase, flow is very similar to that realized in the isothermal run-away collapse phase.
In panel (d), we plotted the structure captured in L16 at $t=0.724236\tau_{\rm ff}$ 
 ($\tau=4.138\times 10^{-3}\tau_{\rm ff}$).
After the central density exceeds $\rho_{\rm C}$,
 another adiabatic core (the second core) is formed (panel e).
By a similar mechanism to form a bipolar outflow, a second outflow is formed around the second core (panel f).
This is a snapshot of L16 at $t=0.724237\tau_{\rm ff}$  ($\tau=4.140\times 10^{-3}\tau_{\rm ff}$).
}

\centering
\hfill (a)\hfill (b)\hfill \\
\hfill (c)\hfill (d)\hfill \\
\hfill (e)\hfill \\
\figcaption{
Mass inflow/outflow rates measured at the boundaries of nested grid systems (eqs.\ref{eqn:mdotin}
 and \ref{eqn:mdotout}). 
Panel (a)-(e) corresponds to models AH1, BH, CH, DH, and EH.
The  horizontal axis represents the time after the core formation.
Levels at which the mass inflow/outflow rates are measured are shown near the respective lines.
We plotted L11 in a solid line, L10 in a dotted line, L9 in a short-dashed line,
 L8 in a long-dashed line, L7 in a short dash-dotted line, and L6 in a long dash-dotted line.
In panel (e) to specify which line represents the outflow, we added brackets to the outflow rates.
Inflow rates are larger than outflow rates for respective levels.
\label{fig9}
}

\centering
\hfill (a)\hfill (b)\hfill \\
\hfill (c)\hfill (d)\hfill \\
\hfill (e)\hfill \\
\figcaption{
Linear momentum outflow rates measured at the upper and lower boundaries of nested grid systems (eqs.\ref{eqn:mvdotout}). 
Panel (a)-(e) corresponds to models AH1, BH, CH, DH, and EH.
The horizontal axis represents the time after the core formation.
Levels at which the linear momentum outflow rates are measured are shown near the respective lines.
We plotted L11 in a solid line, L10 in a dotted line, L9 in a short-dashed line,
 L8 in a long-dashed line, L7 in a short dash-dotted line, and L6 in a long dash-dotted line.
\label{fig10}
}

\end{document}